\documentclass[preprint]{ptephy_v1}

\preprintnumber{XXXX-XXXX} 
\usepackage{arydshln}
\usepackage{subcaption}
\usepackage{caption}
\usepackage{multirow}
\usepackage{arydshln}
\usepackage{amssymb}
\usepackage{amsmath}
\usepackage{url} 
\usepackage{xfrac}
\usepackage{lscape}
\usepackage{booktabs}

\begin{document}
\title{Development of a method to measure trace level of uranium and thorium in scintillation films}


\author[1, *]{K.~Ichimura}
\author[1]{K.~Chiba}
\author[1,4]{Y.~Gando}
\author[1]{H.~Ikeda}
\author[1,5]{Y.~Kishimoto}
\author[1]{M.~Kurasawa}
\author[1]{K.~Nemoto}
\author[2]{A.~Sakaguchi}
\author[2]{Y.~Takaku}
\author[3]{Y.~Sakakieda}

\affil[1]{Research Center for Neutrino Science, Tohoku University, 6-3 Aoba, Aramaki, Aoba-ku, Sendai, Miyagi 980-8578, Japan}
\affil[2]{Institute of Pure and Applied Sciences, University of Tsukuba, 1-1-1 Tennodai, Tsukuba, Ibaraki,
305-8577, Japan}
\affil[3]{Graduate School of Science and Technology, University of Tsukuba, 1-1-1 Tennodai, Tsukuba, Ibaraki,
305-8577, Japan}
\affil[4]{Department of Human Science, Obihiro University of Agriculture and Veterinary Medicine, Inada-cho Nishi 2-11, Obihiro, Hokkaido 080-8555, Japan}
\affil[5]{Kavli Institute for the Physics and Mathematics of the Universe (WPI), The University of Tokyo Institutes for Advanced Study, The University of Tokyo, 5-1-5 Kashiwanoha, Kashiwa, Chiba 277-8582, Japan}
\affil[*]{\rm E-mail: ichimura@awa.tohoku.ac.jp}

\begin{abstract}%
We have established a method to measure picograms-per-gram~(pg~g$^{-1}$) levels of $^{238}$U and $^{232}$Th in scintillation films by combining the dry ashing method and inductively coupled plasma mass spectrometry.
Trace amounts of $^{238}$U and $^{232}$Th were measured in up to 2~g of the scintillation film with almost 100\% collection efficiency.
This paper details the experimental procedure, including the pretreatment of the samples and labware, detection limit of the method, collection efficiencies of $^{238}$U and $^{232}$Th, and measurement of $^{238}$U and $^{232}$Th in a polyethylene naphthalate film.
This method is also applicable to $^{238}$U and $^{232}$Th measurements in other low-background organic materials for rare event search experiments.
\end{abstract}

\subjectindex{H20}

\maketitle

\section{Introduction}
Discovery of neutrinoless double-beta decay (0$\nu\beta\beta$) signals is one of the most important goals in modern particle physics.
This phenomenon would reveal the Majorana nature of the neutrino and can explain the extremely light neutrino mass via a seesaw mechanism~\cite{10.1143/PTP.64.1103, Gell-Mann:1979vob, AKHMEDOV2000215}.
To observe the 0$\nu\beta\beta$ signal, the detector requires many  0$\nu\beta\beta$ candidate nuclei and an ultralow-background (BG) environment.

The KamLAND Zero-Neutrino Double-Beta Decay (KamLAND-Zen) experiment searches for a 0$\nu\beta\beta$ signal by using 745 kg of enriched Xe~\cite{PhysRevLett.130.051801, PhysRevC.107.054612}. 
Approximately 90\% isotopically enriched $^{136}$Xe gas was dissolved in a liquid scintillator at 3~wt\% and stored in an ultralow-BG nylon inner balloon (IB) with a thickness of 25~$\mu$m~\cite{Gando_2021}.
The IB is surrounded by 1~kton of the liquid scintillator contained in an outer balloon with a radius of 13~m. 
The outer liquid scintillator acts as an active shield against the ambient $\gamma$-ray BG.
To date, KamLAND-Zen has set the most stringent upper limit on the Majorana neutrino mass worldwide~\cite{PhysRevLett.130.051801}.
One of the BGs in the current KamLAND-Zen 0$\nu\beta\beta$ search is the $^{214}$Bi $\beta$ decay originating from the IB.
This BG can be identified by delayed coincidence with its progeny nuclide $^{214}$Po.
However, if $^{214}$Po $\alpha$ decay occurs inside the IB nylon film and $\alpha$-rays do not escape from the film, the $\alpha$ signals cannot be detected.
To eliminate such $^{214}$Bi BG, the use of a scintillation film that emits scintillation light from the $^{214}$Po $\alpha$ energy deposited in the film has been proposed.
According to a prior feasibility study~\cite{10.1093/ptep/ptz064}, $^{214}$Bi--$^{214}$Po sequential decay in a polyethylene naphthalate (PEN) scintillation film can be tagged with an efficiency of 99.7\%.
Assuming radioactive equilibrium, the orders of the concentrations of $^{238}$U and $^{232}$Th in the PEN film for KamLAND-Zen requirement should be equal to or less than~10$^{-12}$~g~g$_{\rm PEN}^{-1}$~\cite{10.1093/ptep/ptz064}. 
In this case, the expected BG from the PEN film is comparable to the inevitable $^{8}$B solar-neutrino electron recoil BG.

In addition to the KamLAND-Zen experiment, other experiments such as LEGEND~\cite{Manzanillas_2022} and PICOLON~\cite{10.1093/ptep/ptab020} have the possibility to use ultralow-BG PEN.
Thus, the development of a ultralow-BG PEN and a method for measuring trace amounts of $^{238}$U and $^{232}$Th are indispensable.
Inductively coupled plasma mass spectrometry (ICP-MS) is a method for measuring trace amounts of $^{238}$U and $^{232}$Th.
This highly sensitive mass spectrometry method is widely used in chemistry to measure nuclides in solutions.
A solution with the organic components removed must be prepared in order to measure the elements in organic materials via ICP-MS.
To remove the organic components, the wet ashing method for measuring the amounts of $^{238}$U and $^{232}$Th in the PEN structural component has been reported~\cite{Efremenko_2022}.
In the reported method, approximately 0.2~g of PEN is digested in nitric acid~(HNO$_{3}$) at 250${}^\circ$C and measured using ICP-MS.
A detection limit of aproximately 1 picogram-per-gram of PEN~(pg~g$_{\rm PEN}^{-1}$) has been achieved for both $^{238}$U and $^{232}$Th by using the wet ashing method~\cite{Efremenko_2022}.

In this study, we demonstrated that the dry ashing method, wherein a large amount of PEN is microwaved and ashed for ICP-MS measurement, achieves a method detection limit (MDL) of a few~pg~g$_{\rm PEN}^{-1}$ with a collection efficiency of almost 100\%. Sect.~\ref{sec:procedure} provides details of the method and its performance such as MDL and collection efficiencies of $^{238}$U and $^{232}$Th.
Sect.~\ref{sec:results} presents the results of the PEN film analysis. The conclusions drawn from this study and potential improvements are presented in Sect.~\ref{sec:Conclusion}.

\section{Method and its performance}
\label{sec:procedure}
\begin{figure}[htbp]
\includegraphics[width=\linewidth]
{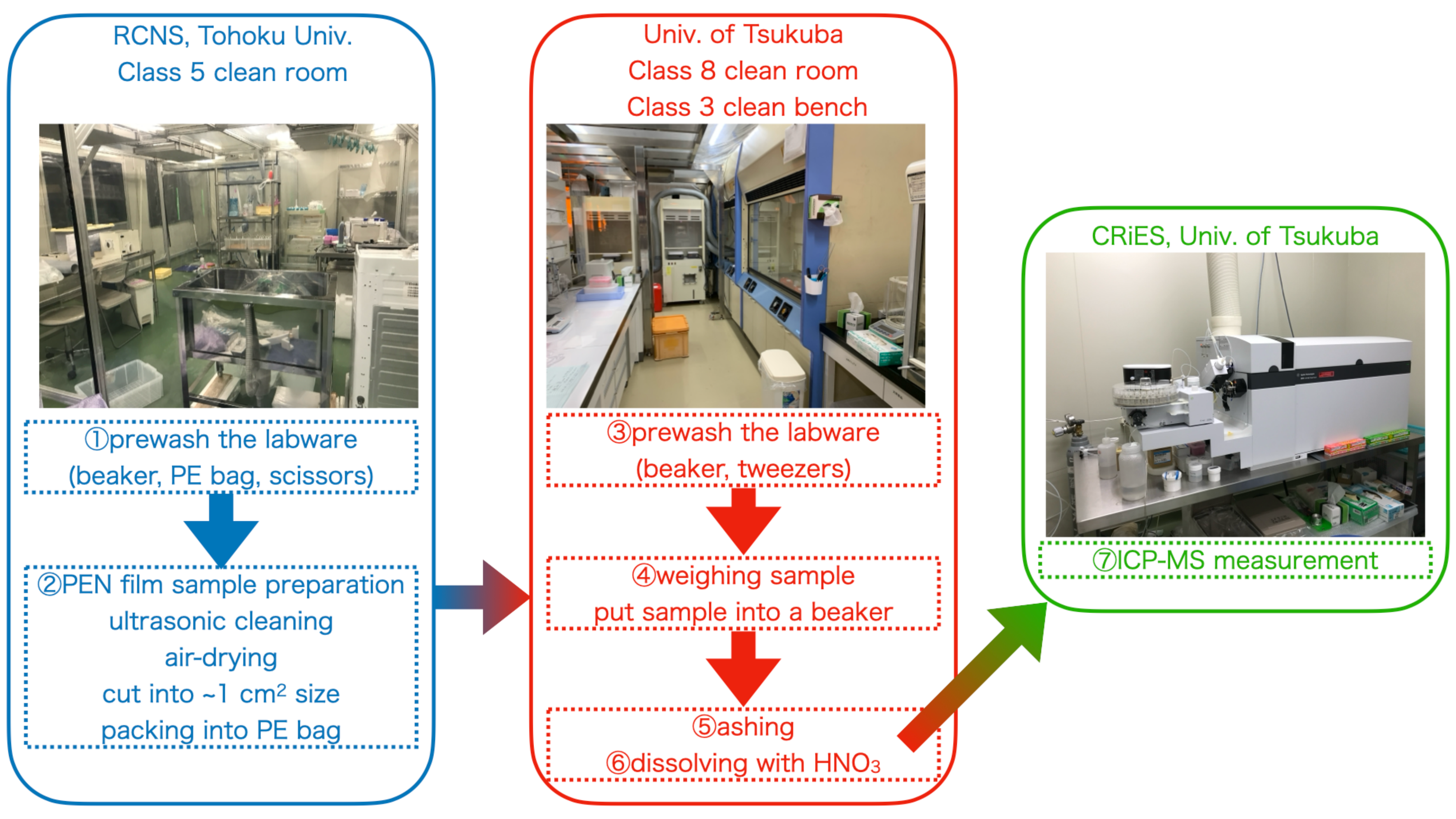}
\caption{schematic view of the entire procedure for measuring of $^{238}$U and $^{232}$Th in a PEN film.}
\label{fig:Diagram}
\end{figure}

Fig.~\ref{fig:Diagram} shows a schematic view of the entire procedure for measuring the concentration of $^{238}$U and $^{232}$Th in a PEN film. Details of each step are presented in this section. 
The ultrapure-grade HNO$_{3}$ used in this study was procured from TAMAPURE-AA-100 (Tama Chemicals Co., Ltd.). 
The standard solution refers to XSTC-331 (SPEX CertiPrep), which originally contained 10~mg~L$^{-1}$ of $^{238}$U and $^{232}$Th.
The ISO 14644-1 standard was used to define the air cleanliness classes for clean rooms and benches. 
We used 1~L and 100~mL quartz beakers made from fused silica.
Further, a highly transparent PEN film manufactured by TEIJIN Ltd. was used in this study.
\subsection{Method for measuring $^{238}$U and $^{232}$Th in the PEN film}
\label{sec:Sec2_1}
\subsubsection{Pre-treatement of the sample and labwares at RCNS, Tohoku University}
Sample preparation, including washing, cutting, and packing of the PEN films, was performed in a Class~5 clean room at the Research Center for Neutrino Science (RCNS), Tohoku University.

\paragraph{Prewash the labware}
To remove contaminations on the surface of the sample and labwares, the 1~L beakers used for ultrasonic cleaning and the polyethylene (PE) zipper bags (UNIPACK\textsuperscript{\textregistered}, SEISANNIPPONSHA, Ltd.) used for transportation were first soaked in diluted ultrapure-grade HNO$_{3}$ for several days and then in ultrapure water for at least one day.
The concentration of ultrapure-grade HNO$_{3}$ for soaking was set to be 0.7~mol~L$^{-1}$ (0.15~mol~L$^{-1}$) for beakers (PE bags).
Finally, these items were rinsed with ultrapure water.
The ceramic scissors used to cut the PEN films were cleaned with ultrapure water in an ultrasonic cleaner for 15~minutes before use.

\paragraph{PEN film sample preparation}
\label{sec:PENPreparation}
Outside the clean room, the PEN films were pre-cut to a size that would fit into a 1~L beaker.
Then, inside the clean room, the PEN films were rinsed with ultrapure water, cleaned with ultrapure water for 10~minutes by using an ultrasonic cleaning machine, and naturally dried immediately below a high-efficiency particulate air (HEPA) filter for at least 24~hours.
After removing the moisture on the PEN films, they were cut into pieces smaller than 1~cm$^{2}$. These pieces were packed in the PE bags and transported to the University of Tsukuba.

\subsubsection{Treatments of the sample and labwares at University of Tsukuba}
\paragraph{Prewash the labware}
\label{sec:TsukubaPreWash}
The 100~mL beakers used for ashing were first soaked in 1.5~mol~L$^{-1}$ electronic-grade HNO$_{3}$ (Kanto Chemical Co. Inc.) at 90${}^\circ$C for at least two~days in a chemical fume hood.
The beakers were soaked in ultrapure water at the same temperature for the same durations.
Finally, the beakers were soaked in a 10-fold diluted high-purity alkaline cleaning solution (TMSC solution; TAMA Chemicals Co., Ltd.) at room temperature until use.
When these beakers were used, they were flushed with ultrapure water, rinsed inside once with 1~mL of 15.2~mol~L$^{-1}$ ultrapure-grade HNO$_{3}$, and washed twice more with 2~mL of 0.15~mol~L$^{-1}$ ultrapure-grade HNO$_{3}$ on a Class~3 clean bench.
They were then cleaned again with ultrapure water and dried naturally on a clean bench.
The tweezers used to weigh the PEN film were cleaned with ultrapure water by using an ultrasonic cleaning machine.

\paragraph{Weighing sample}
The PEN film pieces transported from Tohoku University described in Sect.~\ref{sec:PENPreparation} were placed in prewashed 100~mL quartz beakers while being weighed with an electronic balance on a Class~3 clean bench. The samples were then subjected to the ashing process.

\subsubsection{Ashing and making sample solution}
\label{sec:AshingAndSolution}
\begin{figure}[htbp]
\includegraphics[width=\linewidth]
{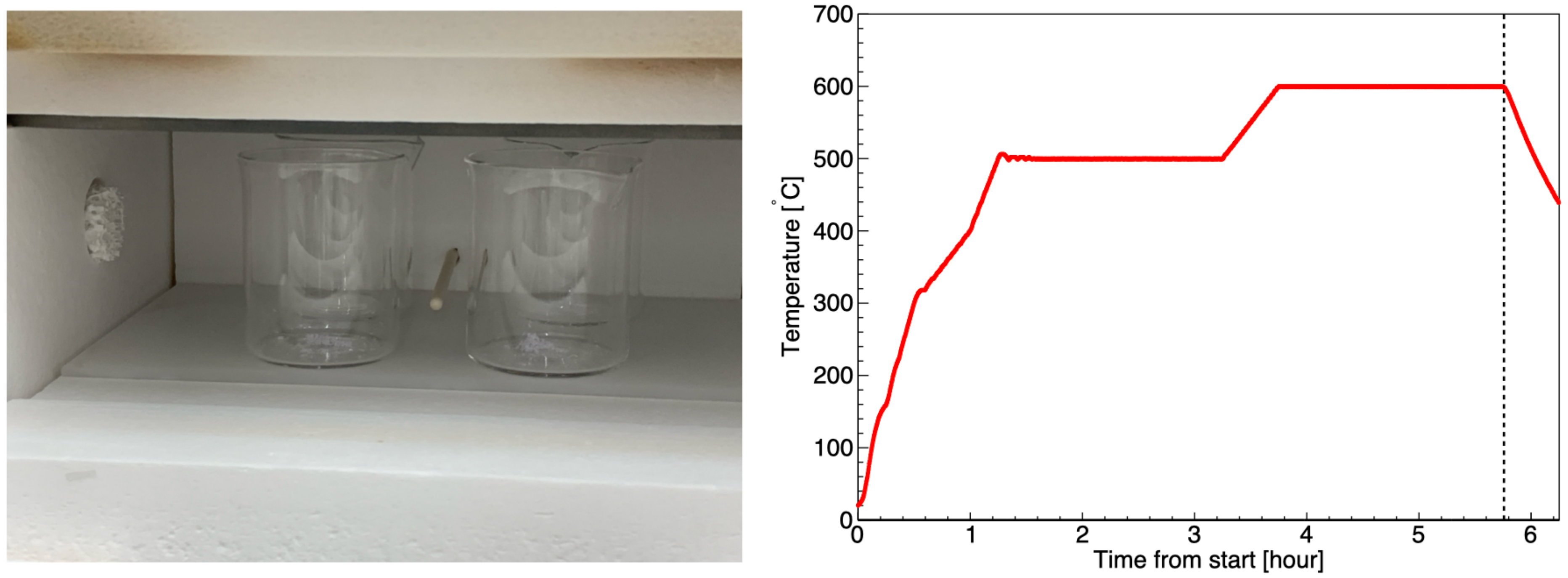}
\caption{(Left) Photograph of the interior of a muffle furnace after ashing. (Right) Time variation of the temperature inside a muffle furnace.
The vertical dotted line indicates the time at which the ashing process is complete.
The system power was turned off at 6.2 hours.}
\label{fig:PYRO}
\end{figure}

The PYRO microwave ashing system manufactured by Milestone Srl. was used for ashing.
The left photograph in Fig.~\ref{fig:PYRO} shows the interior of a PYRO high-purity Al$_{2}$O$_{3}$ muffle furnace.
Four 100~mL quartz beakers were placed in the furnace.
The temporal variation in the temperature inside the furnace was plotted on the right-hand side of Fig.~\ref{fig:PYRO}.
Because the melting point of PEN is approximately 270${}^\circ$C and organic materials tend to become ash at approximately 500--600${}^\circ$C, the temperature was set to increase slowly from 300${}^\circ$C to a higher temperature to prevent sudden boiling.
The temperature in the muffle furnace was maintained at 500${}^\circ$C for two hours and then at 600${}^\circ$C for two hours to completely ash the sample.
PEN films with a mass of 2~g were ashed in 8 hours operation including the cooling time, and the organic component was completely vaporized.
To dissolve the residue after ashing, 1~mL of 15.2~mol~L$^{-1}$ ultrapure-grade HNO$_{3}$ was first added.
Then, 1~mL of 0.15~mol~L$^{-1}$ ultrapure-grade HNO$_{3}$ was dripped onto the sides of the beaker, and the inner surface was washed by shaking the beaker. 
This process was repeated four times to ensure that no residue remained at the bottom or sides of the beaker.
With regard to the solution of ash from the 2~g PEN film used in this study, no visible undissolved residue was observed.
Finally, 5~mL of ultrapure-grade HNO$_{3}$ sample solution ($\approx$ 5~g$_{\rm solution}$),  consisting of 1~mL of 15.2~mol~L$^{-1}$ ultrapure-grade HNO$_{3}$ and 4~mL of 0.15~mol~L$^{-1}$ ultrapure-grade HNO$_{3}$, was transferred to a vial for the ICP-MS measurement.
All of the treatments after asing were performed on a Class~3 clean bench.

\subsubsection{ICP-MS measurement}
In this study, an Agilent 8800 triple quadrupole ICP-MS installed at the Center for Research in Radiation, Isotope and Earth System Science (CRiES), University of Tsukuba was used to measure trace amounts of $^{238}$U and $^{232}$Th in units of (g~g$_{\rm solution}^{-1}$). 
A quartz nebulizer with a sample uptake flow rate of 400 $\mu$L~min$^{-1}$ was used to introduce samples into the torch.
As described in a prior study~\cite{10.1093/ptep/ptad117}, an ion lens called an s-lens (Agilent Technologies) was used to optimize the analysis of low-matrix samples.
To maximize the detection sensitivity of $^{238}$U and $^{232}$Th, ICP-MS instrument parameters such as gas flow rates, torch position, and lens bias were tuned using a 100~ng~g$_{\rm solution}^{-1}$~(100~ppb) $^{238}$U and $^{232}$Th solution, which was obtained following 100-fold dilution of the standard solution.
Each nuclide of interest in the sample was measured thrice with an integration time of three seconds.
Calibration solutions with concentrations of 0.1, 0.5, 1, 5, 10, and 50~pg~g$_{\rm solution}^{-1}$~(ppt) of $^{238}$U and $^{232}$Th were also measured to convert counts per second (CPS) to concentrations (pg~g$^{-1}_{\rm solution}$) for each measurement and sample.
The detection limit (DL) of the ICP-MS was calculated as:
\begin{equation}
\rm{DL}  = \frac{3S_{\rm blank}}{C},
\end{equation}
where $\rm S_{ blank}$ and C are the standard deviations of CPS for the blank solution and slope of the calibration curve, respectively~\cite{10.1093/ptep/ptx145}.
The measurement of an ultrapure water sample indicates that the detection limits of the ICP-MS are 1.1~fg~g$^{-1}_{\rm solution}$~(ppq) for $^{238}$U and 8.1~fg~g$^{-1}_{\rm solution}$ for $^{232}$Th.

\subsection{Detection limit of the method}
\label{sec:MDL}
To evaluate the detection limit of this method, the procedure blank was evaluated for contamination from the beaker and ashing process.
Nine beakers were subjected to the washing, ashing, and dissolution processes, as  described in Sect.~\ref{sec:TsukubaPreWash} and \ref{sec:AshingAndSolution} without the addition of PEN film pieces. Further, $^{238}$U and $^{232}$Th in the solutions were measured using ICP-MS.
The results are summarized in Table~\ref{tab:ProcedureBlankResults}.
It was found that some procedure blanks had relatively high amounts of $^{238}$U and $^{232}$Th (e.g., $^{238}$U in procedure blank \#8 and $^{232}$Th in procedure blank \#1).
Although no clear reason is known, there may be variations in the amount of impurities in the beakers themselves, insufficient washing due to the placement of the beakers during washing, and some chemical species of $^{238}$U and $^{232}$Th that may not be sufficiently removed by washing.
The MDL is set as
\begin{equation}
\rm{MDL}  = \bar{X}_{pb} + 3 S_{pb}
\end{equation}
where $\rm \bar{X}_{pb}$ and $\rm S_{pb}$ are the mean value and standard deviation, respectively, of the procedure blanks, in units of concentration~(g~g$^{-1}_{\rm solution}$). 
The MDL per unit mass of the solution for $^{238}$U and $^{232}$Th were 0.69 and 1.13 pg~g$^{-1}_{\rm solution}$, respectively.
Assuming that the weight of the solution~(g$_{\rm solution}$) is 5~g, and the weight of the PEN film pieces for ashing~(g$_{\rm PEN}$) is 2~g, MDL per unit mass of the PEN film was calculated as 1.7~pg~g$^{-1}_{\rm PEN}$ for $^{238}$U and 2.8~pg~g$^{-1}_{\rm PEN}$ for $^{232}$Th, respectively.
It should be noted that the MDL is low enough to measure $^{238}$U and $^{232}$Th at the levels of a few pg~g$^{-1}_{\rm PEN}$, although the amounts of $^{238}$U and $^{232}$Th vary from procedure blank to procedure blank.

\begin{table}[htbp]
    \centering
    \caption{Results of nine procedure blanks. The average $\bar{X_b}$, standard deviation $S_{b}$ and method detection limit~(MDL) for $^{238}$U and $^{232}$Th were calculated.}
    \begin{tabular}{ccc} \hline
    Blank No. &  $^{238}$U [pg g$^{-1}_{\rm solution}$] & $^{232}$Th [pg g$^{-1}_{\rm solution}$] \\ \hline
   1 & 0.19 & 1.00 \\
   2 & 0.06 & 0.49 \\
   3 & 0.14 & 0.43 \\
   4 & 0.12 & 0.34 \\
   5 & 0.23 & 0.30 \\
   6 & 0.26 & 0.29 \\
   7 & 0.15 & 0.29 \\
   8 & 0.60 & 0.55 \\
   9 & 0.12 & 0.24 \\ \hline
$\bar{X}_{b}$ & 0.21 & 0.44 \\
$S_{b}$ & 0.16 & 0.23 \\
MDL & 0.69 & 1.13 \\ \hline
    \end{tabular}
\label{tab:ProcedureBlankResults}
\end{table}

\subsection{Collection efficiency of this method}
\label{sec:ProcedureRecoveryRate}
\begin{figure}[htbp]
\includegraphics[width=\linewidth]
{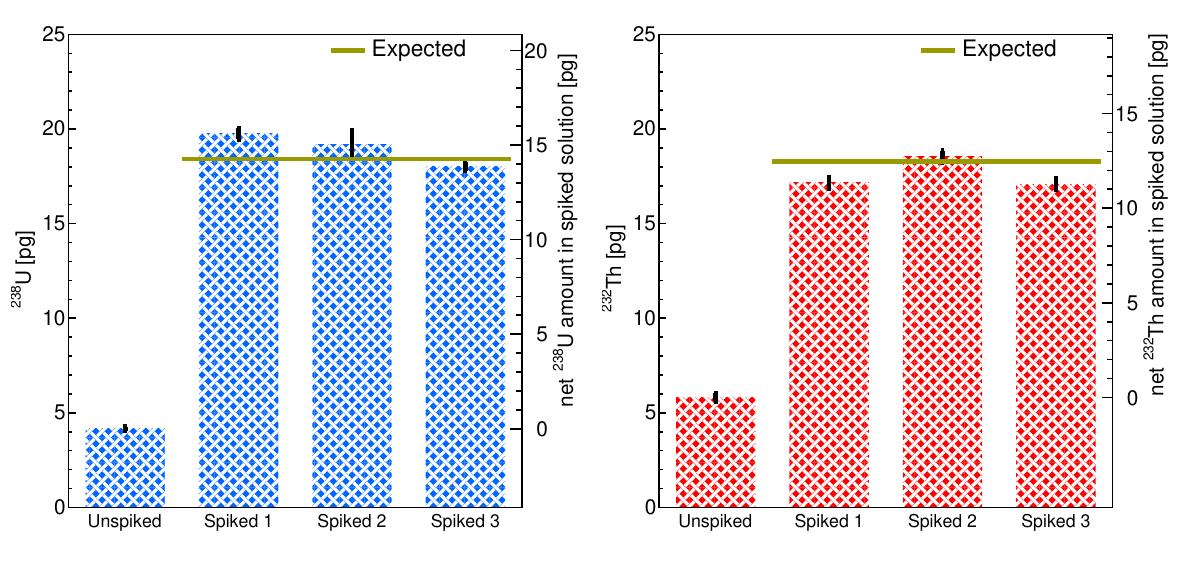}
\caption{Results of the ICP-MS measurement of $^{238}$U (left) and $^{232}$Th (right) for the recovery rate study.
Error bars correspond to the standard deviation of three times measurements for each sample.
The expected line is derived from the sum of the unspiked result and the standard solution measurement, assuming 100\% collection efficiency.}
\label{fig:RecoveryRate}
\end{figure}

The amounts of $^{238}$U and $^{232}$Th that can be collected from the PEN film pieces must be estimated by using the procedure described in Sect.~\ref{sec:Sec2_1}.
Because there are no certified reference materials for PEN, we conducted the spike and recovery test. We used the standard solution to evaluate the collection efficiencies of $^{238}$U and $^{232}$Th.
Three samples of 2~g PEN film pieces spiked with 100~$\mu$L of a standard solution with a concentration of 100 pg~g$^{-1}$ of $^{238}$U and $^{232}$Th (spiked~1,~2,~3), and one sample of 2~g PEN film pieces without a standard solution (Unspiked) underwent the ashing and dissolution processes and were measured by ICP-MS.
Fig.~\ref{fig:RecoveryRate} shows the results for spiked and the unspiked samples. 
The weighted average of the spiked sample results subtracted from the unspiked sample result was found to be 14.6$\pm$0.5~pg for $^{238}$U and 11.8$\pm$0.5~pg for $^{232}$Th.
In contrast, the expected amounts of $^{238}$U and $^{232}$Th from  standard solution measurements are 14.3$\pm$0.2~pg and 12.5$\pm$0.4~pg, respectively.
On the basis of these values, the collection efficiencies regarded as recovery rates were found to be 102.4$\pm$3.6\% for $^{238}$U and 94.3$\pm$3.8\% for $^{232}$Th.
This nearly 100\% recovery rate can be used as a reference value indicating that the ashing and dissolution process can collect trace amounts of $^{238}$U and $^{232}$Th without loss.

\subsection{Evaluation of the possible contamination from PE bag and Class~5 clean room}
\label{sec:RCNSContamination}
To study the maximum contamination from PE bags and Class~5 cleanroom, eight prewashed PE bags were prepared.
The inner surfaces of the seven PE bags were washed with 10~mL of 0.15~mol~L$^{-1}$ ultrapure-grade HNO$_{3}$ for each bag.
The remaining bag was left in a Class~5 cleanroom for one day and then washed in a similar manner.
After these washed, the HNO$_{3}$ solutions were analyzed by ICP-MS.
The results of the seven PE bags measurements were 0.043$\pm$0.012~pg~g$^{-1}_{\rm solution}$ for $^{238}$U and $<$0.048~pg~g$^{-1}_{\rm solution}$ for $^{232}$Th.
The error value of $\pm$0.012 for the $^{238}$U measurement corresponded to the standard deviation of seven measurements, and the result of the $^{232}$Th measurement was below the detection limit.
On the basis of this result and the assumption of the weight of the solution (5~g$_{\rm solution}$) and sample (2~g$_{\rm PEN}$) described in Sect.~\ref{sec:MDL}, the maximum contaminations from the PE bag were calculated to be 0.11$\pm$0.03~pg~g$^{-1}_{\rm PEN}$ for $^{238}$U and $<$0.12~pg~g$^{-1}_{\rm PEN}$ for $^{232}$Th.
The results for the remaining PE bag left in the clean room for one day were 0.13$\pm$0.01~pg~g$^{-1}_{\rm PEN}$ for $^{238}$U and $<$0.12~pg~g$^{-1}_{\rm PEN}$ for $^{232}$Th.
Here, the error value of $\pm$0.01 for the $^{238}$U measurement corresponds to the standard deviation of three replicate measurements.
Thus, contamination from the PE bag and the environment in the Class~5 clean room was found to be negligible for $^{238}$U and $^{232}$Th measurements at a detection pg~g$^{-1}_{\rm PEN}$ levels.

\section{Results of the PEN film analysis}
\label{sec:results}
\begin{figure}[htbp]
\includegraphics[width=\linewidth]
{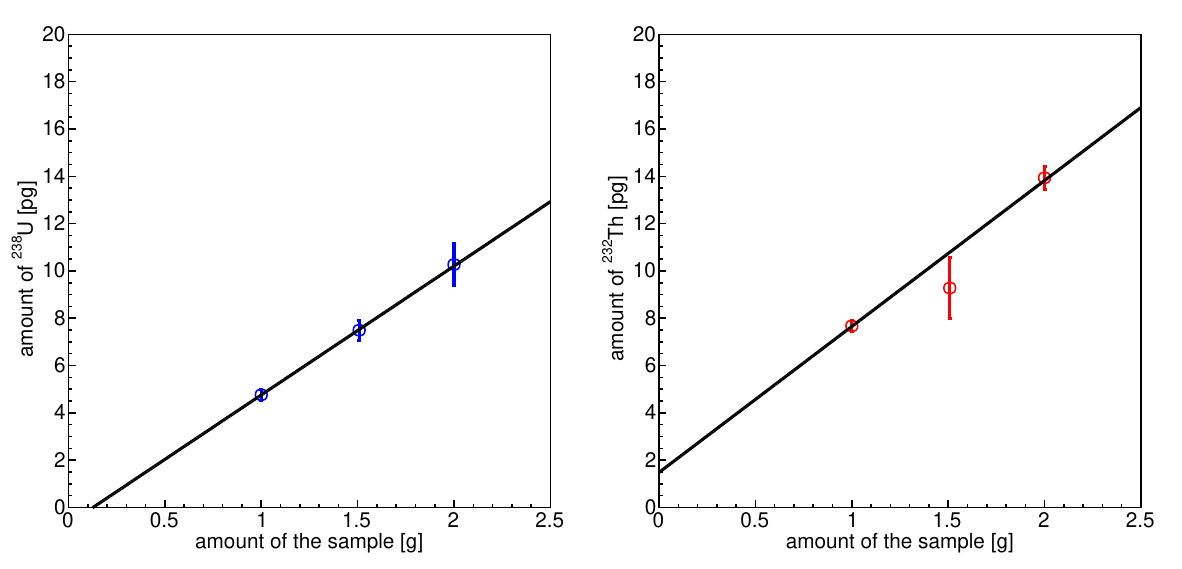}
\caption{Results of the ICP-MS measurements of $^{238}$U (left) and $^{232}$Th (right) with three different masses of the PEN film.
Error bars correspond to the standard deviation of three sample measurements for each mass.
The solid line shows the fitted result and this slope corresponds to the amount of  $^{238}$U and $^{232}$Th in the sample.}
\label{fig:PENRI}
\end{figure}

We measured amounts of $^{238}$U and $^{232}$Th in the PEN film using Patterson's plot method~\cite{PattersonPlot, JA9940901385}. 
This method utilizes the linear correlation between sample mass and the amounts of $^{238}$U and $^{232}$Th.
Three samples of three different masses (1.0, 1.5, and 2.0~g) of PEN film pieces were used to determine concentrations of the $^{238}$U and $^{232}$Th.
Fig.~\ref{fig:PENRI} shows the relationship between the sample mass and amounts of $^{238}$U and $^{232}$Th.
Here, the error bars represent the standard deviation of the measurement results for the three samples.
The size of error bars for the result of 2.0~g ($^{238}$U) and 1.5~g ($^{232}$Th) is relatively large, possibly due to the variation in procedure blanks mentioned in Sect.~\ref{sec:MDL}. 
However, these errors are approximately 15\%, which is small enough for a measurement at the level of a few~pg~g$^{-1}_{\rm PEN}$.
The values of $^{238}$U and $^{232}$Th were extracted from slopes of the linear functions, as shown in Fig.~\ref{fig:PENRI} and found to be 5.4$\pm$0.7~pg~g$^{-1}_{\rm PEN}$ for $^{238}$U and 6.2$\pm$0.5~pg~g$^{-1}_{\rm PEN}$ for $^{232}$Th.
This indicates that the impurities in the PEN film satisfied the KamLAND-Zen requirement.
This also demonstrates that our method can measure a few~pg~g$^{-1}_{\rm PEN}$ levels of $^{238}$U and $^{232}$Th in PEN films.

\section{Conclusion and possible improvements} 
\label{sec:Conclusion}
We established a method to measure trace amounts of $^{238}$U and $^{232}$Th in PEN films by combining the dry-ashing method and ICP-MS with an almost 100\% recovery rate.
PEN film pieces with a mass of 2~g were ashed, and the detection limit of the method was found to be a~few~pg~g$^{-1}_{\rm PEN}$ from the procedure blank results.
Contamination from the PE bag and the environment in the Class~5 clean room was found to be negligible at the level of a few~pg~g$^{-1}_{\rm PEN}$.
We measured $^{238}$U and $^{232}$Th in the PEN films at levels of a~few~pg~g$^{-1}_{\rm PEN}$ by using Patterson's plot method.
This method can be applied to other organic materials such as 1,4-Bis(2-methylstyryl)-benzene (Bis-MSB) wavelength shifter.
If a PEN film with higher radiopurity is developed in the near future, 
the MDL must be accordingly improved to measure its concentration by using this method.
We plan to improve the cleanliness of the work areas related to Sect.~\ref{sec:TsukubaPreWash} and~\ref{sec:AshingAndSolution}.
Contamination is expected to be reduced by increasing the number of HEPA filter units and introducing a Class~5 clean draft to use as the prewash as described in Sect.~\ref{sec:TsukubaPreWash} and high-purity synthetic quartz beakers.
The sensitivity of ICP-MS can be improved by a factor of 5 by introducing an Aridus3$^{\rm TM}$ solvent-removal module (Teledyne CETAC Technologies)~\cite{10.1093/ptep/ptad117}.
These improvements would facilitate the detection of $^{238}$U and $^{232}$Th at much lower concentrations in organic materials. 

\section*{Acknowledgment}
This research was supported by Japan Society for the Promotion of Science (JSPS) KAKENHI Grantin-Aid for Scientific Research on Innovative Areas, with grant numbers 19H05803, 21H01104, and 21H01105. This work was also supported by the Environmental Radioactivity Research Network Center (ERAN) F-22-16 and P-23-09.

\bibliographystyle{ptephy2}
\bibliography{Reference}

\end{document}